# The thin ring wing as a means of flow improvement upstream of a propeller.

Vladimir Sluchak

*ABSTRACT*

*There are numerous devices currently known with the purpose to reduce the irregularity of the flow upstream of the propeller and to decrease by that means the propeller-induced vibration and noise. Many of these devices are wing-shaped vortex-generators that affect the flow with their induced (i.e. passive) longitudinal vortices. The subject of the paper is the use of a ring-shaped wing as a highly effective passive vortex-generator which allows to control the flow closer to the most charged sections of propeller blades. The problem of a thin ring-shaped wing with irregular (asymmetric) geometry in the irregular steady flow has been solved in linear approach and the intensity of the induced longitudinal vortices as a function of the irregularity of the flow and the geometry of the ring wing has been estimated using that solution. Experiments in the towing tank showing good concordance with the theoretical model confirmed the effectiveness of such a device. Some additional advantages of a ring-shaped wing incorporated into the construction of stabilizers are considered.*

As the level of a propeller noise often needs to be lowered certain kinds of "flow improvement" upstream of a propeller make sense and is widely used. The currently admitted model of the flow filtration by a propeller allows to formulate the following flow improvement targets [1]:

to decrease a high frequency (with continuos spectrum) component of the propeller noise the rough characteristic of the flow irregularity forward of the propeller is to be decreased - the difference $\Delta$ between maximum and minimum of the circular distribution $V(\varphi,r)$ of the velocity at end sections of blades;

to decrease discrete spectrum components of the vibration and the revolution sound of the propeller it is enough to decrease amplitudes and sums of amplitudes $|a_{qz}|$; $|a_{qz\pm1}|$ ; $|b_{qz}|$ ; $|b_{qz\pm1}|$ of harmonics (member of the corresponding Fourier series) of $V(\varphi,r)$ at the most charged sections of propeller blades. Here: $z$ is the number of propeller blades ; $q$ - digit ; $a$ , $b$ - cosine and sine harmonic amplitudes.

The harmonics with numbers more than $z$ are very hard to control because they practically never dominate in $V(\varphi , r)$ making some hundredths of $\Delta$. So the second part of the flow improvement problem is much more delicate then the first one and actually has never been approached effectively enough. Usually the revolution sound of the propeller is optimized by some expensive way of propeller design which in particular implies the increase of blades numbers. The described flow improvement problem is principally solved in two ways: the optimization of the vessel architecture (the shape and dimensions of hull , sail, stabilizers, covers etc. ) and the usage of special flow improvement devices (later - FID).

For different kinds of vessels (submarines and submersibles - at most) a ring-shaped wing placed forward of the propeller around the tale of a hull presents a very effective combination of these two ways of flow improvement. Such a wing architecturally and functionally perfectly fits into the tale -plane or stabilizers allowing to decrease stabilizers dimensions crucial for the magnitude of $\Delta$. You may consider the following architecture for a submarine with a traditional pushing propeller as a $\Delta$- minimizing one : an axis - symmetrical hull , no sail , stabilizers as a ring - shaped wing on relatively narrow pillars (unlike conventional cross - shaped wing ) . Such an architecture accepted as an ideal - all the real architectural types might be put in some order of deviation from it.

Let's traverse that architectural sequence trying to keep the same low level of flow irregularity forward of a propeller: the more an architecture would deviate from the ideal one (for instance - the larger the sail would be ) the more flow improving work would be required. Incorporating the ring - shaped wing within the architecture allows to do that work very effective way .

The ring affects the flow by its passively generated longitudinal vortices. The intensity of an elementary vortex running down from the ring is proportional to the derivative $\partial(vb\varepsilon)/\partial\varphi$, where v is the velocity of the flow on the front of the ring , b - the chord of the ring , $\varepsilon$ - local angle of attack composed of a constructive angle of the ring wing and the angle between the flow direction and the axes of the ring. As your can see from that formula the vortex shroud running down from a ring wing is sensitive to the irregularity of the flow upstream : the more intense is the irregularity the greater is the intensity of generated vortices. The vortices intensity may be controlled precisely at any sector of the propeller's disk with variation of the ring geometry which the simplest case is circular variation of the chord. The generated vortices work aft of the ring in the boundary layer of a hull principally the same way as for the most of other passive vortex - generating FIDs : they induce traversal flows that redistribute the layers with different longitudinal velocity (fig.1).

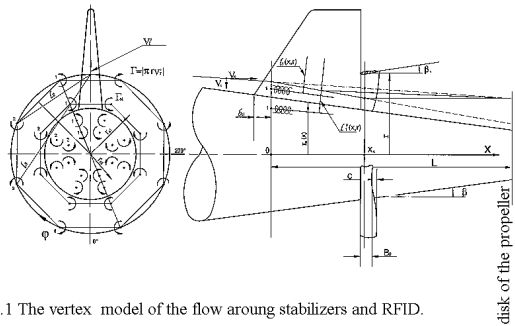

Fig.1 The vertex model of the flow aroung stabilizers and RFID.

Assuming the wing be relatively narrow and thin and deferring the estimation of limits of such an assumption until some time later , let us represent the ring wing with one flat circular attached vortex with circulation $\Gamma(\varphi)$ and cylindrical (with radius R) shroud of free vortices with intensity $(1/R)\,\partial\Gamma/\partial\varphi$. The flow forward of the ring is considered ideal and given by its three components of the velocity on the front edge of the ring wing. The profile of the wing is considered symmetrical , the distance b/4 from the attached vortex to the front edge of the wing - small enough to disregard the induction of the attached vortex on the front edge of the ring wing. Within the limits of such a simple model the equation for the circulation $\Gamma(\varphi)$ is derived on the assumption that the Chapligin-Jukovsky postulate is met in every meridian section of the ring wing . Introducing the adjustment $\varepsilon_i(\varphi)$ to the local angle of attack $\varepsilon(\varphi)$ which allows for the effect of a vortex shroud we have:

$$\Gamma(\varphi)= \pi b(\varphi)V(\varphi)[\varepsilon(\varphi) - \varepsilon_i(\varphi)] \qquad (1)$$

where:

$$V(\varphi) = \sqrt{V_r^2(\varphi) + V_x^2(\varphi)}$$
$$\varepsilon_i(\varphi) = V_{r_i}(\varphi,R)/V(\varphi) \;;$$

Placing the statement for radial velocity induced by the free semi-infinite cylindrical vortex shroud on the axis of the circular attached vortex:

$$V_{r_i}(\varphi, R) = \frac{1}{8\pi R} \int_0^{2\pi} \frac{\partial \Gamma}{\partial \varphi'} \operatorname{ctg} \frac{\varphi' - \varphi}{2} d\varphi' \qquad (2)$$

, we get the following equation for the circulation $\Gamma(\varphi)$ :

$$\Gamma(\varphi) = \pi \tilde{\varepsilon}(\varphi) - \frac{b(\varphi)}{8R} \int_0^{2\pi} \frac{\partial \Gamma}{\partial \varphi'} \operatorname{ctg} \frac{\varphi' - \varphi}{2} d\varphi' \qquad (3)$$

where :

$\tilde{\varepsilon}(\varphi) = V(\varphi)b(\varphi)\varepsilon(\varphi)$ and $b(\varphi)$ is the chord.

The equation (3) is solved by means of expansion of $\Gamma$ , b and $\tilde{\varepsilon}$ into Fourier series , while the following integrals provide the key to the solution :

$$\int_0^{2\pi} e^{in\varphi'} \operatorname{ctg} \frac{\varphi' - \varphi}{2} d\varphi' = -2\pi i e^{in\varphi} \qquad (4)$$

, where n - natural and $i = \sqrt{-1}$;

Substituting $\tilde{\varepsilon}$, $\Gamma$ and b with their Fourier series:

$$\begin{cases} \tilde{\varepsilon} = \tilde{\varepsilon}_0 + \sum_{k=1}^{\infty}(A_k \cos k\varphi + B_k \sin k\varphi); \\ \Gamma(\varphi) = \Gamma_0 + \sum_{k=1}^{\infty}(C_k \cos k\varphi + D_k \sin k\varphi); \\ b(\varphi) = b_0 + \sum_{k=1}^{\infty}(E_k \cos k\varphi + F_k \sin k\varphi); \end{cases}$$

and using (4) we get the infinite system of algebraic equations linear and fully defined for Fourier coefficients of $\Gamma$. For the case of constant or almost constant chord ($b(\varphi) \approx b_0$) that system comes down to the following:

$$C_k = \pi A_k/(1+\pi k/2\lambda); \quad D_k = \pi B_k/(1+\pi k/2\lambda) \quad (5)$$

This solution - if to make a quick test of its correctness - allows to derive the known formulae for coefficients of lifting force and the induced drag of the ring wing in the steady regular flow:

$$\left. \begin{array}{l} C_Y^\alpha = \dfrac{\pi}{1+\dfrac{\pi}{2\lambda}} \\ C_X = \dfrac{(C_Y^\alpha)^2}{2\lambda} \end{array} \right\} \qquad (6)$$

The adjustment factor $1 - (e^{-\lambda})/2$ improves these formulae for small $\lambda$ (fig. 2):

$$C_Y^\alpha = (1 - \frac{e^{-\lambda}}{2}) \frac{\pi}{1+\dfrac{\pi}{2\lambda}} \qquad (7)$$

Even if installed exclusively to work as flow improvement device with its longitudinal vortices as a main factor the ring wing would partly substitute the initial stabilizer (usually - a cross-shaped wing). Therefore it is interesting to compare coefficients of lifting force for ring-shaped and cross-shaped wings.
In the table below S stands for the area, D - for the ring diameter, L - for the cross-shaped wing span and Y - for the lifting force, "+" denotes cross-shaped wing.

| Conditions | Ratio $C_y / C_y+$ |
|---|---|
| (S= S+, $\lambda = \lambda$+) | $(\lambda + 2)/(\lambda + \pi/2)$ |
| (S = S+, D= L) | $(\lambda + \pi)/(\lambda + \pi/2)$ |
| (D = L, Y = Y+) | $(2\lambda/\pi + 1)/(2\lambda/\pi - \pi/2\lambda)$ |

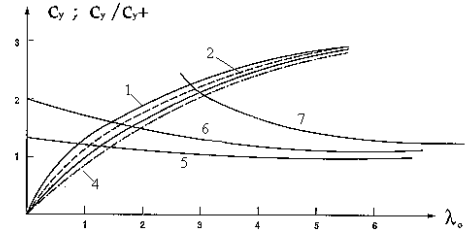

Fig. 2a The coefficient of lifting force Cy of the ring wing (1--4) and its comparison to the one of cross-shaped wing (5–7).
Denoted: 1 -- formula (6), 2 -- Weissinger, 3 -- formula (7), 4 -- empirical, 5 -- $\lambda = \lambda+$, $S_0 = S+$, 6 -- for $l=b$, and $S_0 = S+$, 7 -- for $l=b$ and $Y=Y+$

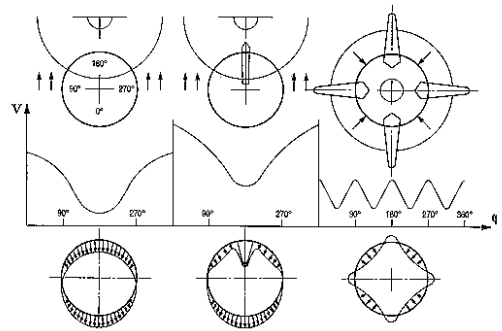

Fig. 2b Radial velocities induced by the ring wing in different flows.

It may be observed from this comparison that the substitution of cross-shaped stabilizers by ring-shaped ones allows to win in stabilizers span, while keeping the same hydrodynamic quality. In particular it is easy to make the span less than the diameter of a hull.
Now let us analyze the work of a flow improving ring on the simplified example of a flow forward of the propeller of a single shaft submarine (fig. 1). In this case the distribution of longitudinal velocities on the front edge of a ring wing is approximately described with following formulae:

$$\left. \begin{array}{l} V = \Delta_v \cos\varphi + V_0 \\ \varepsilon = \Delta_\varepsilon \cos\varphi + \varepsilon_0 \end{array} \right\} \qquad (8)$$

Taking into account that the outlook of circular distributions of the flow velocity is defined by longitudinal vortices generated by cross-shaped stabilizers, let us define the ring wing chord as follows:

$$b = \Delta_b \cos 4\varphi + b_0 \qquad (9)$$

where $b_0$ - the average chord, $\Delta_b$ - the amplitude of the chord. Using (9), (8), (5) and (2) we get the following expression for the radial velocity induced by semi-infinite cylindrical shroud of vortices on the core of an attached vortex (fig. 2):

$$\overline{V}_{r_i}(\varphi) = f_4 \cos 4\varphi + f_8 \cos 8\varphi + f_{12} \cos 12\varphi \qquad (10)$$

Where:

$$\overline{V}_{r_i}(\varphi) = V_{r_i}(\varphi)/V_0;$$

$$f_4 = \frac{\Delta_\varepsilon (\frac{3}{4}\overline{\Delta}_v \overline{\Delta}_\varepsilon + 1) + \varepsilon_0 (\overline{\Delta}_v \overline{\Delta}_b + \overline{\Delta}_v + \overline{\Delta}_b)}{1 + \frac{\lambda}{2\pi}};$$

$$f_8 = \frac{\overline{\Delta}_\varepsilon (\overline{\Delta}_v + \overline{\Delta}_b)}{2(1 + \frac{\lambda}{4\pi})};$$

$$f_{12} = \frac{\overline{\Delta}_v \overline{\Delta}_\varepsilon \overline{\Delta}_b}{4(1 + \frac{\lambda}{6\pi})}; \qquad \overline{\Delta}_v = \frac{\Delta_v}{V_o}; \overline{\Delta}_b = \frac{\Delta_b}{b_o};$$

It may be observed from that expression that the velocities induced by the vortex shroud of a ring wing:
a) have directions opposite to the ones of velocities induced by stabilizers' vortices,
b) are proportional to the irregularity of the flow forward of the ring,
c) are amenable to control by means of variations of the ring geometry.

When the ring is in the slanting flow (as if the vessel were maneuvering or moving straight but with an angle of attack), the induction of the vortex shroud leads to the partial restoration of the axis symmetry of the flow aft of the ring forward of the propeller. In that case the distribution of velocities on the front edge of the ring is approximated with the formula:

$$\varepsilon = b_0 V_0 (\varepsilon_0 + \alpha \cos\varphi),$$

where $\alpha$ - is the angle between the axis of a ring and a flow direction.

For the distribution of radial velocities on the thread of the attached vortex we have:

$$V_{ri} = \alpha \cos\varphi/(1 + 2\lambda/\pi) \qquad (11)$$

The formulae (10) and (11) show two important features of such a FID that make it different from all the others known for axis-symmetrical hulls (single shaft submarines for instance):

a) automatism of the flow improvement effect,
b) the ability to reduce so-called hull irregularity of the flow that is observed when moving with angle or maneuvering.
The third important feature of the ring-based FID (later on - RFID) emerges directly from its architecture: the ability to affect particular sectors of the flow forward of the propeller.
That last feature may be particularly useful for non axis-symmetrical vessels. As we mentioned before it implies the use of a theory of relations between the "outlook" (the set of some rough characteristics) of the circular distribution of the velocity and its Fourier series.

Having observed main features of a RFID we might set the design procedure of such a device as fully experimental: sequential adjustments of the ring geometry during model tests in towing tank. But there are some design steps that may be pre-calculated at least for axis-symmetrical vessels. These steps are:
1. The optimization of ratios between the major dimensions of a RFID and the local diameter of the hull from the condition of maximum induction of vortex shroud when moving at an angle of attack.
To solve that problem the hull is imitated with a dipole in the plane of an attached vortex of a ring wing which drives the problem down to the equation:

$$\partial A(r, b)/\partial r = 0,$$

where:

$$A(r, b) = \frac{1 - (r_0/r)^2}{1 + \frac{2\lambda}{\pi}};$$

2. The optimization of the dimensions of RFID from the condition of smooth (non separating) flow in the channel formed by a ring and a hull.
The limit ratio between in and out sections area is used here:
$S_{out}/S_{in} = 1.2$, that insures the non-separating flow inside of a diffuser.
3. The estimation of a maximum opening angle of a RFID meeting the condition of non-separating flow on the under-pressured (sucking) side of the wing. The ring is devised into segments with angular dimensions defined by maximums of the function $\partial \Gamma/\partial \varphi$. The empirical formula for the critical angle of attack may be used here:

$$\varepsilon_{cr} \approx 30°/\sqrt{\lambda_p},$$

where $\lambda_p$ - the elongation of a segment.
4. The estimation of the circular variation of the chord of a RFID from the condition of a maximum

effect on circular distributions of the velocities on the given radius of a propeller's disk.

Formula (9) is accepted for a chord. The axis - symmetrical boundary layer is pre-calculated using any known method (for instance - [2] ).

Following assumptions are enforced :

a) the cross - shaped symmetrical stabilizers is the sole assembly of parts jutting out of the hull ,

b) the stabilizers affect the flow only with their head vortices,

c) the threads of head vortices of stabilizers remain in meridian planes laying under angles $\varphi = \pi(2k +1)/8$ to the vertical plane.

The value of $\Delta_b$ is found from the condition of zero shift of the intersection of the given flow line
(laying in the plane of a stabilizer) with the plane of the propeller's disk (fig.1 ).

These calculations allow to draw some diagrams to estimate the geometry of a RFID.

The final design of an RFID for single shaft submarine is to be done on basis of model experiments with RFID. As of two shaft submarines and other non axis - symmetrical vessels - the series of experiments with stepwise adjustments of the construction is practically the only way to design RFID.

Passing to the experimental part of the RFID study it is difficult to refrain of making the suggestion that a smart incorporation of a ring wing in tail assemblies of vessels almost inevitably leads to the decrease of flow irregularity and necessarily - propeller noise and vibration. The impressive result of such a design is observed on torpedoes with total substitution of cross-shaped stabilizers by a ring -shaped wing (fig. 3) which is obviously the case when the whole flow improving effect comes from placing the main surface of stabilizers out of the flow running into the propeller's disk. There may by several design solutions of that kind proposed in all of which the vortex shroud of a ring wing does not play any essential role in flow improvement forward of the propeller.

As it has been already noticed experiments with RFID are needed not only to test the theory and the final effect: the main part of the RFID design is supposed to be done as a series of adjustments based on model tests results. When the substantial decrease of the $\Delta$ is achieved and with not too many of experimental adjustments you may want to skip all the intermediate estimations and get right to the sound calculation with a high probability of a success . But, whatever be the decrease of the $\Delta$ you are never sure that it leads to the decrease of the particular harmonic $a_i$ unless you have handy some proven formula of the

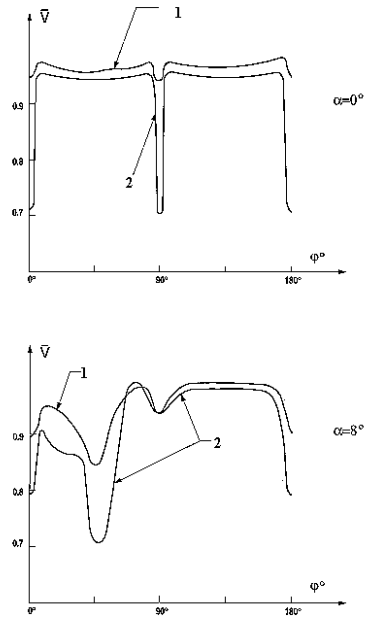

Fig3. The result of cross-shaped stabilizers substitution by a ring wing on a torpedo moving with the angle of attack.
1 - ring wing, 2 - cross-shaped stabilizers.

kind $f_1(\Delta) > f_2(a_i)$ (provided as well that Fourier series of the initial velocity distribution $V(\varphi)$ have been received ) .

Practically all the controllable characteristics of the circular velocity distributions $V(\varphi)$ at different radii of the propeller are the "rough" ones comparing to the real aim of the control - the harmonics of $V(\varphi)$.

The $\Delta$ may be considered as the roughest one , all the others like for instance :
maximum derivative of $V(\varphi)$ , local extremums,

or the full deviation $\int_0^{2\pi} |dV(\varphi)|$ - as more detailed ones. The very control of $V(\varphi)$ is in a sense a process of putting certain not fully defined picks on that curve within some sectors. To choose optimal sectors for the picks and to predict their effect on the harmonics of the curve $V(\varphi)$ the theory is required. Such a theory , kind of Fuzzy Fourier Analysis , has been set [3] using the following :

a) theorems of Fourier Coefficients estimations ( the simplest one $|a_i| < 2\Delta/\pi$ ),

b) some features of symmetry of Fourier Coefficients,

c) the concept of the ideal curve (the curve $V_{ideal}(\varphi)$

that does not contain harmonics whose suppression is a goal of the control),

d) the concept of a limiting $\delta$-wide band (the area between two curves defined as $V_{ideal}(\varphi) + \delta/2$ and $V_{ideal}(\varphi) - \delta/2$ ), such that the fall of a certain harmonic amplitude in the certain predefined interval of values is guaranteed if the curve $V(\varphi)$ will remain within the band along the whole interval $[0, 2\pi]$.

Using the described theory we can formulate the practical and reachable goal of $V(\varphi)$ control during the model test as follows: to put $V(\varphi)$ within the predefined band at the predefined ideal curve $V_{ideal}(\varphi)$ (fig. 4), which is actually the goal of the RFID design.

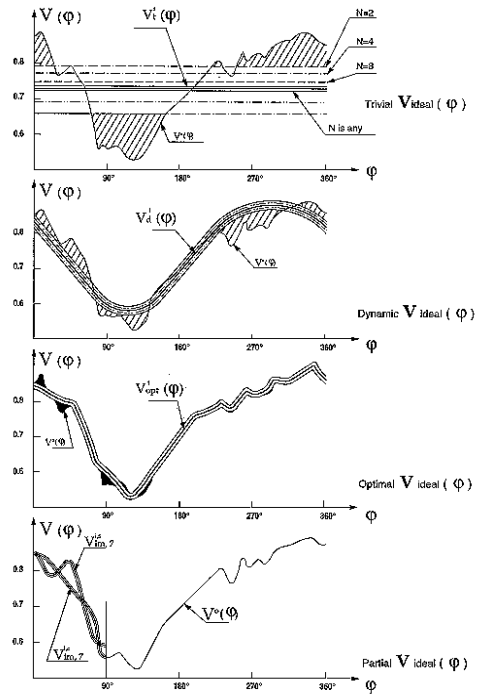

Fig.4 Ideal curves and limiting bands.

The model test with RFID have been carried out on long models (about 20 feet length) in a big towing tank of Krilov Scientific Institute of Hydrodynamics in St. Petersburg (Russia). The calculations of the revolution sound have been performed with the original method and computer program developed in the same Institute (Bavin et al.). The inputs for the program are: the field of flow velocities $V(\varphi,r)$ forward of the propeller, the propeller's detailed geometry and revolution's rate.

To facilitate the empirical design procedure for two shaft submarines it has been found convenient to make models of a RFID with 10°-wide changeable rear winglets(fig.5).

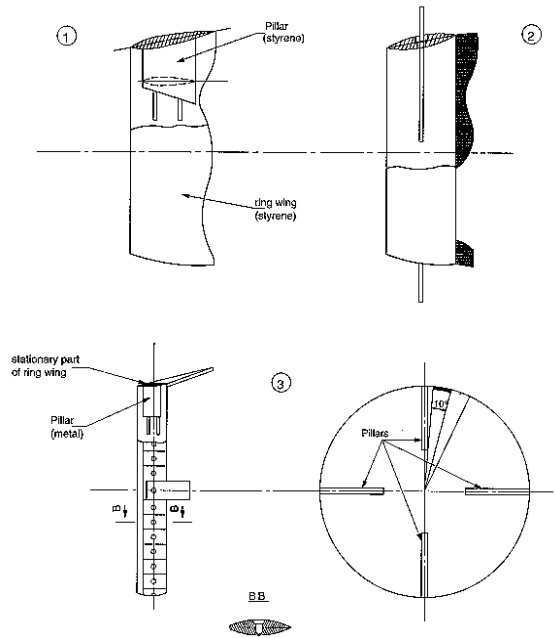

Fig.5 RFID models tested in a towing tank.

As you can see from tests results for one shaft submarine with RFID (fig. 6, 7) the flow improving effect is persistent, the $\Delta$ may be lowered by as much as 60% and the revolution sound - by 6~10 db . The flow improving effect extends to maneuvering which is unique: no other devices are known that could automatically decrease the so-called hull irregularity of the flow. The tests have shown also that in case of small clearance between a hull and the RFID, the diffuser effect may considerably slow the flow down due to a pre-separational conditions. It is possible to use that effect for the improvement of the flow but practically it would be too expensive because of the drag resistance increase. As for two shaft submarines: trying to suppress the $\Delta$ with RFID is hopelessly expensive. A more delicate approach is needed here requiring full use of ideal curves with $\delta$-wide bands and RFID mdel design with changeable rear winglets (fig.4 , fig. 5). The RFID design results shown on fig 8 have been received with reasonable number of towing cart runs.

The design method tested on the models of two shaft submarine is fully applicable to the surface ships.

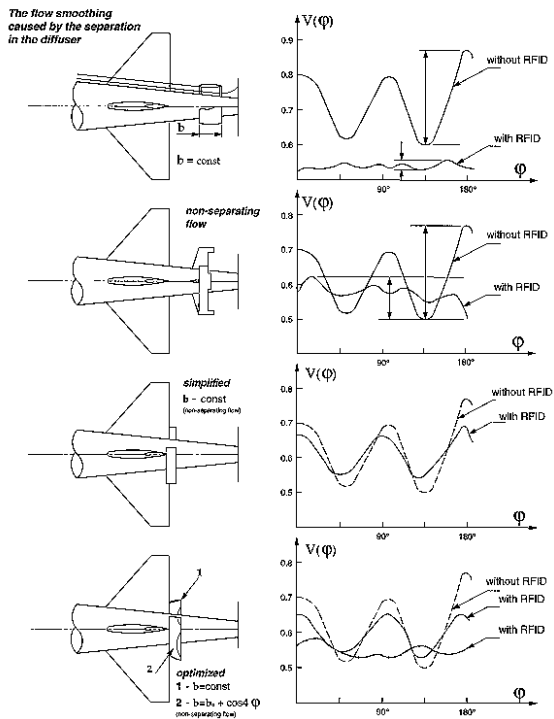

Fig.6 The stability of a RFID effect.

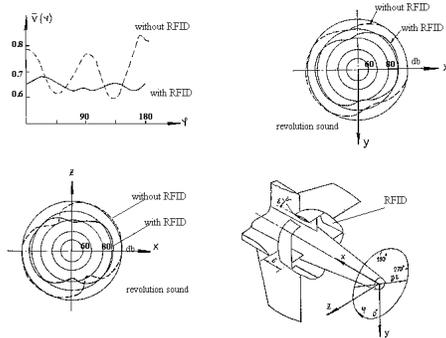

Fig.7 The effect of a RFID on a single shaft submarine (model tests and computations).

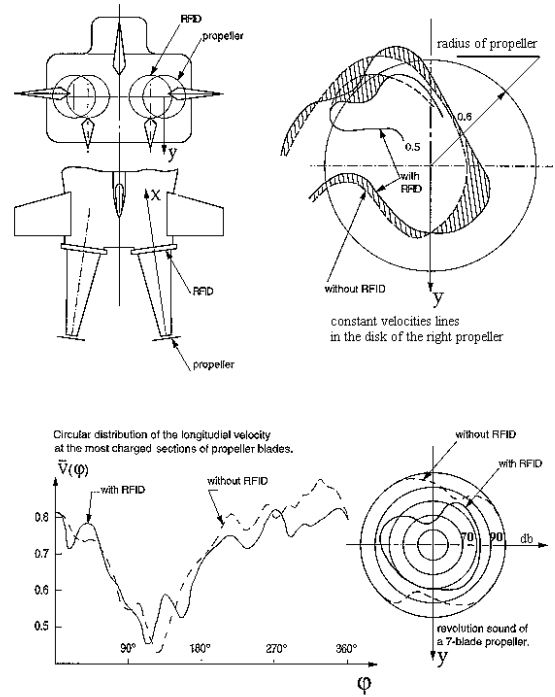

Fig.8 The effect of a RFID on two shaft submarine (model tests and computations).

## Referencies

1. Bavin V.F., Zavadovsky Y.N., Levkovsky Y.L., Mishkevich V.G. Marine Propellers. Modern Methods of Calculation. Sudostroenie Publ. House, Leningrad, 1983 (in Russian; translation into English is made by UK Ministry of Defense, 1991).
2. Larsson L., Broberg L., Keun-Jae Kim, Dao-Hua Zhang A Method for Resistance and Flow Prediction in Ship Design. SNAME Transactions, vol.98, 1990, pp.495-535.
3. Sluchak V., "Method of control of velocity distribution harmonics forward of a Propeller", The Shipbuilding Industry, RUMB, ISS, 6s, 1986, Leningrad. (in Russian)